\documentclass[11pt,twoside]{article}

\usepackage{asp2006}
\usepackage{epsf}
\usepackage{lscape}
\usepackage{epstopdf}
\usepackage[dvips]{graphicx}

\begin{document}
\title{Realistic Numerical Modeling of Solar Magnetoconvection and Oscillations  }   
\author{Irina Kitiashvili$^1$, Laetitia Jacoutot$^1$,
Alexander Kosovichev$^2$, \\
Alan Wray$^3$,
Nagi Mansour$^3$}   
\affil{$^1$Center for Turbulence Research, Stanford University, $^2$HEPL, Stanford University,$^3$NASA Ames Research Center}    

\begin{abstract} 
We have developed 3D, compressible, non-linear radiative MHD simulations to study the influence of the magnetic field of various strength and geometry on the turbulent convective cells and on the excitation mechanisms of the acoustic oscillations. The results reveal substantial changes of the granulation structure with increased magnetic field, and a frequency-dependent reduction in the oscillation power. These simulation results reproduce the enhanced high-frequency acoustic emission observed at the boundaries of active region ("acoustic halo" phenomenon). In the presence of inclined magnetic field the solar convection develops filamentary structure with flows concentrated along magnetic filaments, and also exhibits behavior of running magnetoconvective waves, resembling recent observations of the sunspot penumbra dynamics from Hinode/SOT.
\end{abstract}


\section{Introduction}
 Realistic numerical simulations pioneered by \citet{Stein_2001} have provided an important insight into the structure and dynamics of solar convection and the excitation mechanism of oscillations on the Sun and solar-type stars. The simulations  showed that the acoustic modes by the work of turbulent pressure and nonadiabatic pressure  fluctuations \citep{Stein_2004}. The numerical simulation must rely on sub-grid scale models of turbulence. \citet{jacoutot2008a} investigated various turbulence models and showed that the dynamic model of \citet{moin} provides the best agreement with solar observations. The implementation and details of our MHD code are presented by \citet{jacoutot2008b}.
In this paper, we include magnetic field of the various strength and inclination in the realistic simulations and investigate changes in the physical properties of convection and oscillations. In particular, we investigate the effect of magnetic field on the convection and oscillation spectra,
and study the dynamics of convection in regions of strong highly inclined magnetic field. The simulation results are important for interpretation of high-resolution data from the Solar Optical Telescope \citep{Tsuneta2008} of Hinode \citep{Kosugi2007}.

\section{Granulation structure in vertical magnetic field}
\begin{figure}[t]
\begin{center}
 \includegraphics[width=\textwidth]{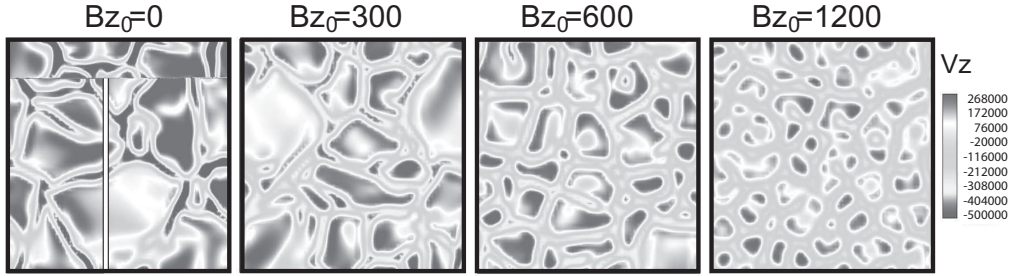}
\caption{Vertical velocity ($cm/s$)  distributions at the visible surface for different initial vertical magnetic fields. The size of the box is 6 Mm.
\label{TVzBz}}
\end{center}
\end{figure}

Figure \ref{TVzBz} shows temperature, vertical velocity, and vertical magnetic field distributions at the visible surface for different magnetic fields.
We observe that the size of granules decreases as the initial magnetic field increases. Without magnetic field the mean size of granules is about 2 Mm, and it is less than 0.75 Mm for 1200 Gauss. In addition, the temperature in the granules becomes higher as the initial magnetic field increases. We can also note that the down-flow in the intergranular lanes is weaker for high magnetic fields.  We can also notice that the magnetic field is swept into the intergranular lanes although the magnetic field is seeded uniformly. This characteristic of solar magnetoconvection confirms previous simulation results \citep[e.g.][]{Stein_2002}.

\section{Oscillation spectrum of acoustic modes in magnetic regions}
\begin{figure}[b]
\begin{center}
 \includegraphics[width=\textwidth]{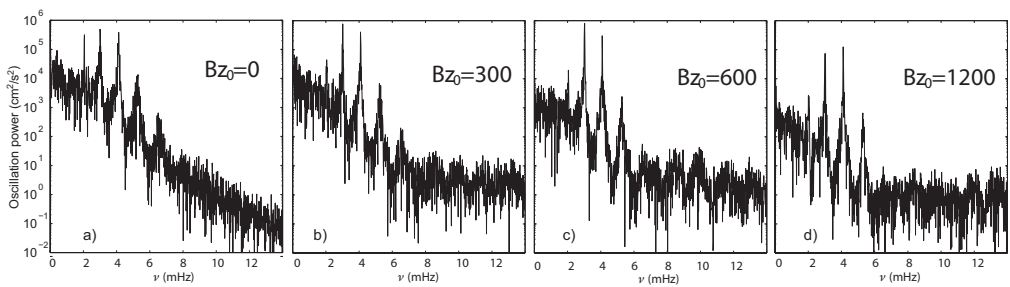}
\caption{Oscillation power spectra of the vertical velocity at the visible surface for different initial vertical magnetic fields.
\label{P_Bz_4panel}}
\end{center}
\end{figure}

We calculate the oscillation power spectra of acoustic
modes of large horizontal wavelength by horizontally averaging the
vertical velocity and Fourier-transforming in time.
These results shown in Fig.~\ref{P_Bz_4panel}
are obtained from the simulations of 60 hours of the solar time using
instantaneous snapshots saved every 30 seconds.
Five oscillation modes can be clearly seen as sharp peaks in the power spectrum of the vertical velocity oscillations at the solar surface. The resonant frequencies supported by the
computational box are 2.07, 3.03, 4.10, 5.23, 6.52 mHz. In addition, several broad high-frequency peaks at 6-12 mHz corresponding to pseudo-modes \citep{kumar1991} can be identified.

The power distribution is shifted towards higher frequencies with the increase of the field strength. It is particularly interesting that the amplitude of the pseudo-modes increases with field strength and reaches maximum at $B_{z0}=600$ G. This may explain the effect of enhanced high-frequency emission (''acoustic halo") around active regions \citep[e.g.][]{jain2002}. The enhanced emission at frequencies 5-7 mHz appears at the boundaries of active regions where magnetic field is moderate. This corresponds to the simulations results: the pseudo-mode amplitude is high for 600 G field and diminishes at 1200 G.

The simulations also show the enhanced spectral power of the convective background at high frequencies for models with magnetic field, forming plateaux at $\nu > 6$ mHz. This leads to the idea that the acoustic halos are caused by enhanced high-frequency turbulent convective motions in the presence of moderate magnetic field. This is consistent with the decreased granular size in magnetic regions. The smaller scale convection naturally has higher frequencies and, thus, generates more higher frequency acoustic waves than convection without magnetic field. When the field is very strong the sound generation decreases because of suppression of convective motions of all scales. This  explains why the acoustic halos are observed in regions of moderate magnetic field strength at the boundaries of active regions.

\section{Effects of inclined magnetic field on convection}
\begin{figure}[t]
\begin{center}
 \includegraphics[width=\textwidth]{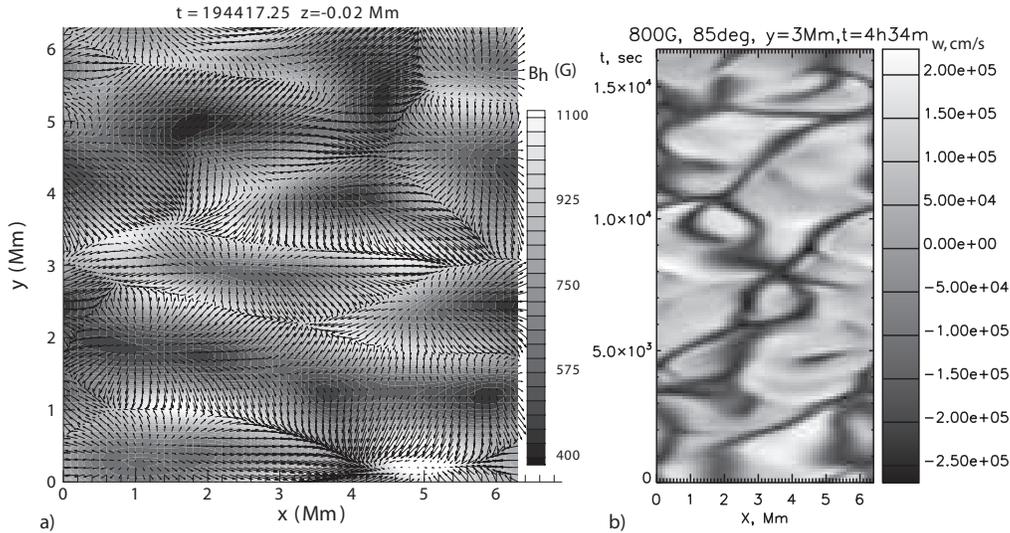}
\caption{a) The horizontal magnetic field strength and velocity field near the surface, showing the filamentary structure of magnetoconvection in the simulations with the initial 800 G magnetic field inclined 85 degrees to the vertical; b) time-distance, $t-x$, propagation diagram for the vertical velocity at the photosphere, showing traveling convection elements.
\label{inclined_field}}
\end{center}
\end{figure}

In addition to the simulations with the initial vertical magnetic field we have carried out simulations of solar convection in presence of inclined magnetic field. The mean magnetic field inclination is maintained at the top and bottom boundary conditions along with the total magnetic flux. The results presented in Fig.~\ref{inclined_field} show that in the presence of inclined field the magnetoconvection pattern is moving in the direction of inclination similar to the running convective wave found in the previous MHD simulations \citep[e.g.][]{Hurlburt1996}. In the realistic 3D case the magnetoconvection develops
filamentary structures (Fig.~\ref{inclined_field}a), and a rapidly moving convection pattern (Fig.~\ref{inclined_field}b). The characteristic speed depends on the field strength and the inclination angle. For a typical case, $B_0=800$ G and $85^\circ$ inclination angle, the speed is about 1--1.2 km/s. The horizontal flow speed reaches 4 km/s. This corresponds to the observed Evershed flow structure and speed, and supports the interpretation of the Evershed effect as a consequence of the thermal convection under the strong inclined magnetic field of sunspot penumbra, from the Hinode/SOT observations \citep{Ichimoto2009}.

\section{Conclusion}
 The realistic simulations of solar convection and oscillation in the presence of magnetic field  reproduce several phenomena observed in solar active regions.
In particular, the results confirm that the spatial scale of granulation substantially decreases with the magnetic field strength. Magnetic field is swept in the intergranular lanes, and the vertical downdraft motions in these lanes are suppressed. This results is a decrease in the excitation power. The oscillation power in the presence of magnetic field is shifted towards higher frequencies, also increasing the amplitude of pseudo-modes above the acoustic cut-off frequency. At a moderate field strength of $\sim 600$ G the power of the high-frequency oscillations reaches a maximum. This corresponds to the phenomenon of ''acoustic halo" observed in the range of 5-7 mHz at the boundaries of active regions.
In the presence of inclined magnetic field the solar convection develops filamentary structure with flows concentrated along the magnetic filaments, and also exhibits behavior of running magnetoconvective waves, resembling recent observations of the sunspot penumbra dynamics from Hinode/SOT presented at this Workshop by \citet{Ichimoto2009}.

\end{document}